\documentclass{sf2a-conf2019}
\usepackage{graphicx}
\usepackage{hyperref}
\usepackage[]{natbib}  
\usepackage{epstopdf}

\def\BibTeX{{\rm B\kern-.05em{\sc i\kern-.025em b}\kern-.08em
    T\kern-.1667em\lower.7ex\hbox{E}\kern-.125emX}}
\bibpunct{(}{)}{;}{a}{}{,}  

\newcommand{\refcre}{C12}

\providecommand{\teff}{\ensuremath{T_{\rm eff}}}
\providecommand{\numax}{\ensuremath{\nu_{\rm max}}}

\providecommand{\fbol}{\ensuremath{F_{\rm bol}}}
\providecommand{\lum}{\ensuremath{L_{\star}}}
\providecommand{\rad}{\ensuremath{R_{\star}}}

\providecommand{\logg}{\ensuremath{\log g}}

\begin{document}

\TitreGlobal{SF2A 2019}


\title{Sun-like Oscillations in the Population II giant HD\,122563}

\runningtitle{Asteroseismology of HD\,122563}

\author{O. Creevey}\address{Universit\'e C\^ote d'Azur, Observatoire de la C\^ote d'Azur, CNRS, Laboratoire Lagrange, France}
\author{F. Th\'ev\'enin$^1$}
\author{F. Grundahl}\address{Stellar Astrophysics Centre, Department of Physics and Astronomy, Aarhus University, Ny Munkegade 120, 8000 Aarhus C, Denmark} 
\author{E. Corsaro}\address{INAF - Osservatorio Astrofisico di Catania, via S. Sofia 78, 95123-I, Catania, Italy}
\author{M. F. Andersen$^2$}
\author{V. Antoci$^2$}
\author{L. Bigot$^1$}
\author{R. Collet$^2$}
\author{P. L. Pall\'e}\address{Instituto de Astrofisica de Canarias, 38205 La Laguna, Tenerife, Spain}
\author{B. Pichon$^1$}
\author{D. Salabert}\address{Previous affil: IRFU, CEA, Université Paris-Saclay, F-91191, Gif-sur-Yvette, France} 

\setcounter{page}{237}


\maketitle


\begin{abstract}
We have been monitoring the metal-poor Population II giant, HD 122563, for radial velocity variations since 2016 using the SONG telescope on Tenerife. We have detected the global seismic quantity \numax\ which provides information related to the stellar parameters. By combining these data with complementary data, we derive a new precise surface gravity, radius and distance to the star. Our results are corroborated by using the parallax from Gaia DR2. We present these results and some of their implications. 
\end{abstract}



\section{Introduction}
{HD\,122563} ($V$=6.2 mag, 14$^{\rm h}$02$^{\rm m}$~31.8$^{\rm s}$, +09$^{\circ}$41${'}$09.95${"}$) is a bright metal-poor [M/H]~=~--2.4 giant star. 
As such it can be observed using many independent methods. 
It was first referenced in the literature over 50 years ago \citep{pagel1963}, however, despite its brightness and interest as a prototype for similar giants in the Galactic halo, today we still debate some of its most fundamental stellar parameters \citep[for example][and references therein]{cre12b,cas14,iva18,collet18}.

Analyses presented in \cite{cre12b} (C12 hereafter, their figure 4) and \cite{cre14} (their figure 3) clearly indicated discrepancies between observations, interpretation and models.  
The former indicated that standard evolutionary tracks fail to reproduce 
the observed position of this star in the HR diagramme and unreasonable assumptions in some
tunable parameters are needed. 
The latter showed a comparison of interferometric with spectroscopic analyses of this star's stellar parameters which hinted towards a potential problem in \logg.

Given the current discrepancies along with the fact that this star is a benchmark for 
distant metal-poor giants, we applied to observe this star using the radial velocity instrument
on the Hertzsprung telescope in Tenerife in order to detect oscillations and provide a fresh
perspective on this star.

In this work we describe the radial velocity observations of this star (Sect.~\ref{sec:observations}),
 along with an interpretation using the asteroseismic scaling relation for \logg\ (Sect.~\ref{sec:analysis}). We use the most recent data from Gaia DR2 to test our analysis, and after some brief comparisons, we summarise our conclusions in Sect.~\ref{sec:conclusions}.
  
\section{Observations\label{sec:observations}}

\subsection{ New observations}

\begin{figure}[ht!]
 \centering
 \includegraphics[width=0.48\textwidth,clip]{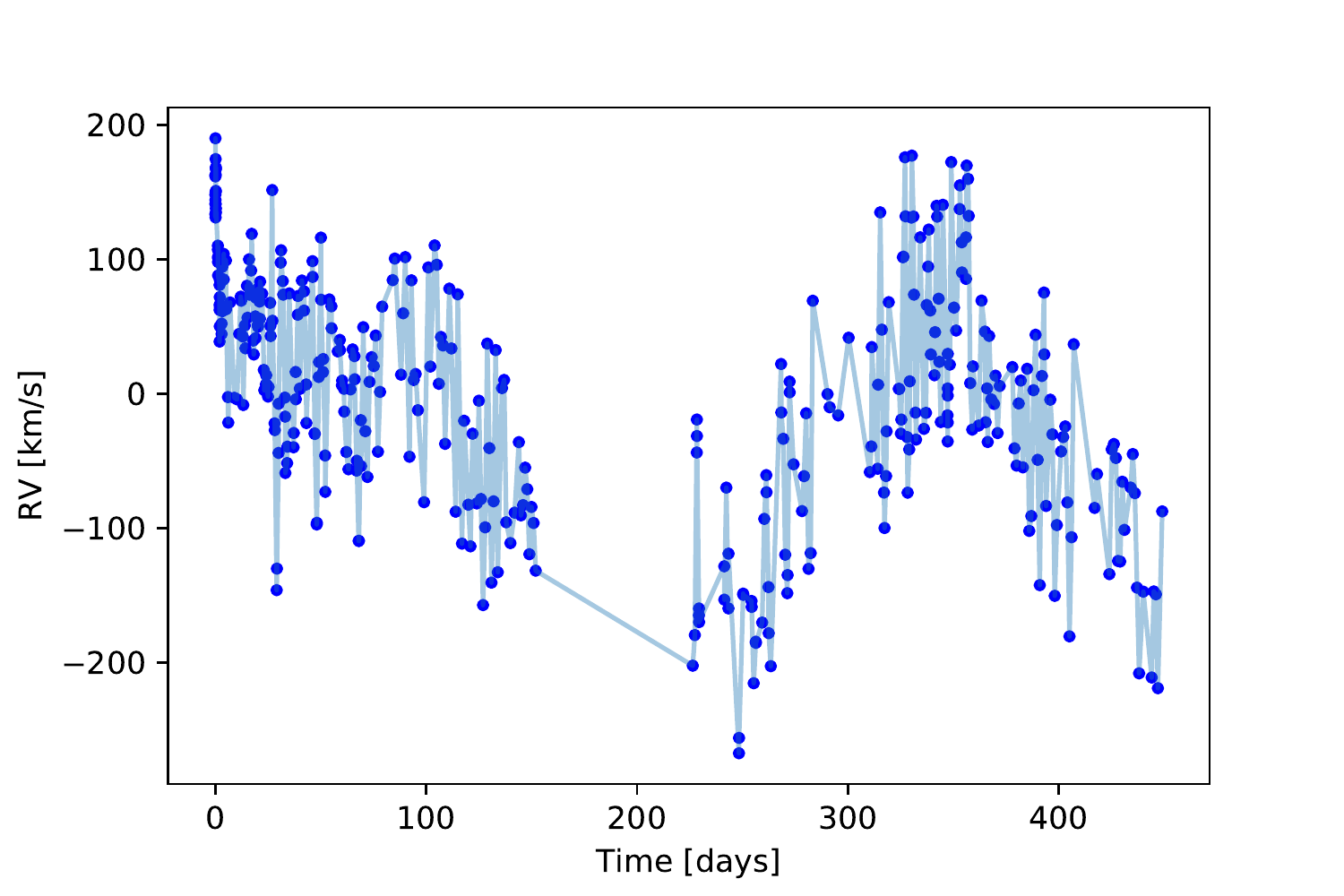}      
 \includegraphics[width=0.48\textwidth,clip]{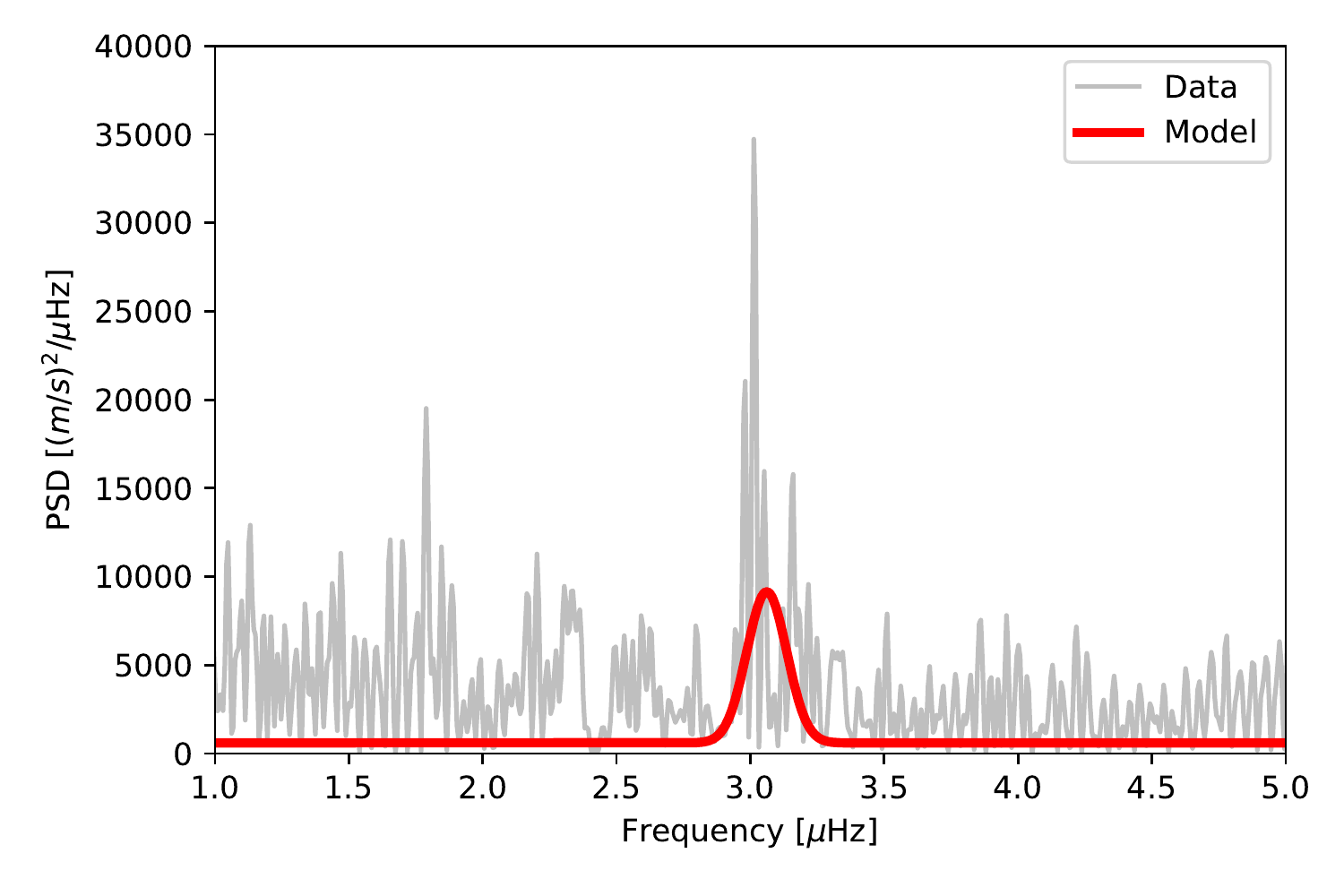}      
  \caption{{\bf Left:} Time series radial velocities of HD\,122563 for the first year of observations.  We have subtracted the mean radial velocity and the first observation point is at time T0 = 2457509.38158 Julian days. {\bf Right:} Frequency spectrum of the time series (grey) with a model for the power excess overplotted in red.}
  \label{Creevey:fig1}
\end{figure}

We obtained time series radial velocity observations with the 1-m Hertzsprung
SONG telescope from April 2016 to December 2017.  The Hertzsprung telescope is a node 
of the Stellar Observations Network Group (SONG) located at the Observatorio del Teide.
All observations were obtained
using an iodine cell for precise wavelength calibration.  A spectral resolution 
of 80\,000 and an exposure time of 900s was used throughout.  The data were reduced using the standard SONG pipeline \citep{anderson2014,grundahl2017}. 
The radial velocity (RV) time series of 387 data points is presented in Figure~\ref{Creevey:fig1}, left panel and  the typical uncertainty on the RV was found to be in the 11-14m/s range. 

We used the \textsc{Diamonds} Bayesian Inference tool \citep{Corsaro14} to model the power spectral density (PSD) of the star. 
The PSD and the best-fit model are shown on the right panel of Fig.~\ref{Creevey:fig1}, and incorporates a flat noise component, {two Harvey-like profiles to account for granulation-driven signal,} and a Gaussian envelope to model the oscillation power excess \citep{Corsaro15}. A clear excess of power due to the
oscillations is detected at 3~$\mu$Hz, this is referred to as \numax.
We used the marginal distribution of \numax\ from this analysis in our subsequent analysis.

\subsection{ Literature observations}
We use the \teff\ of HD\,122563 from  \refcre.
This is in agreement with that derived by \citet{iva18} using independent interferometric measurements, \cite{cas14} using the infra-red flux menthod, and \cite{heiter15} who provide a recommended value based on a compilation of spectroscopic measurements, see \cite{creevey2019}.
As we would like to propagate all of the information from the observations, we use
the reported bolometric flux \fbol\ and angular diameter $\theta$, where extinction $A_V = 0.01$ mag was assumed.

\begin{table}
\begin{center}
\begin{tabular}{lcrcllcrclccc}
\hline
\hline
\multicolumn{5}{c}{Observations}&
\multicolumn{5}{c}{Derived Parameters}\\
\hline
$\theta^1$ & [mas]& 0.940 &$\pm$& 0.011 & \logg & [dex] & 1.39 &$\pm$ &0.01\\
\fbol$^1$ &[erg/cm/s] & 13.16& $\pm$ &0.36 & \logg$_{\rm V17}$ & [dex] & 1.42 &$\pm$ &0.01\\
\teff$^1$ &[K]& 4598& $\pm$ &41 & $d$ & [pc] & 305 &$\pm$& 10 \\
\numax\ &[$\mu$Hz] & 3.07 &$\pm$& 0.05 &$d_{\rm V17}$ & [pc] & 296& $\pm$& 9 \\
$\pi_{\rm GDR2}^2$ & [mass] & 3.444 &$\pm$ &0.063 & $d_{\rm GDR2}$ & [pc] & 290& $\pm$& 5 \\
\hline\hline
\end{tabular}
\caption{Properties derived from this work, except for those from $^1$\refcre\ and $^2$\cite{gaia2018a}.}
\end{center}
\end{table}

\section{Analysis \label{sec:analysis}}
\subsection{ Surface gravity and distance from asteroseismology}
The surface gravity of a star can be derived using the so-called {\it asteroseismic scaling relation} and is given by
\begin{equation}
\frac{\numax}{\nu_{\rm max\odot}} = f_{\nu_{\rm max}}\frac{g}{g_\odot} 
\sqrt{ \frac{T_{\rm eff\odot}}{\teff} } 
\label{eqn:numax}
\end{equation}
where  $f_{\nu_{\rm max}}=1$ and \numax$_{\odot}$ = 3050 $\mu$Hz in the classic form \citep{kb95} and 
$f_{\nu_{\rm max}} \ne 1$ where corrections are proposed. 
By using the observational data ($\theta$,\fbol,\numax) along with the solar values of 
$\log g_{\odot} = 4.438 $, \teff$_{\odot}$ = 5772 K, we performed Monte-Carlo like simulations to derive the stellar parameters (\logg, \teff), using the methods described in \cite{creevey2019}.
By using a prior on the mass between [0.80,0.90] we additionally derived the radius \rad\ and distance $d$ (and consequently the luminosity \lum). 
In figure~\ref{Creevey:fig2} the blue contour lines indicate the density distributions of \logg\ and $d$ from these results.  
The green contours show the same results but by setting $f_{\nu_{\rm max}} = 
(\mu/\mu_{\odot})^{1/2} (\Gamma_1/\Gamma_{1\odot})^{1/2}$, where $\mu$ and 
$\Gamma$ denote respectively
the mean molecular weight and the adiabatic exponent, see \cite{viani2017} (V17 hereon) and references therein.
The derived \lum\ and \teff\ are also shown as the blue error box on the right panel, and for comparison we also show the values from C12 as the grey box.
We indicate a vector in blue which represents the relative change in the position if we impose 
$A_V = 0.08$ mag \citep{stilism}. 

\subsection{ Distance and surface gravity from Gaia DR2}

\begin{figure}[ht!]
 \centering
 \includegraphics[width=0.48\textwidth,clip]{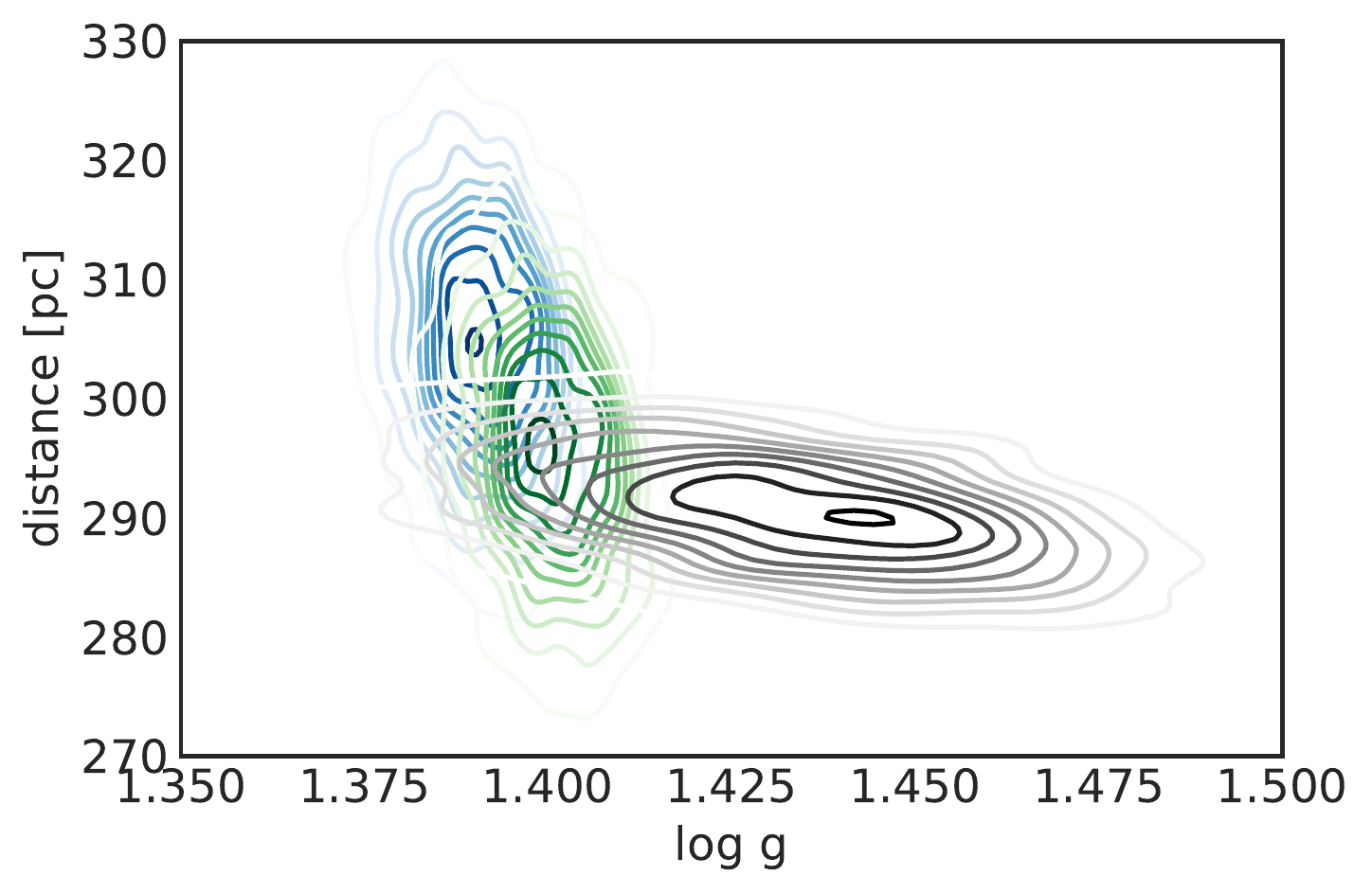}      
 \includegraphics[width=0.48\textwidth,clip]{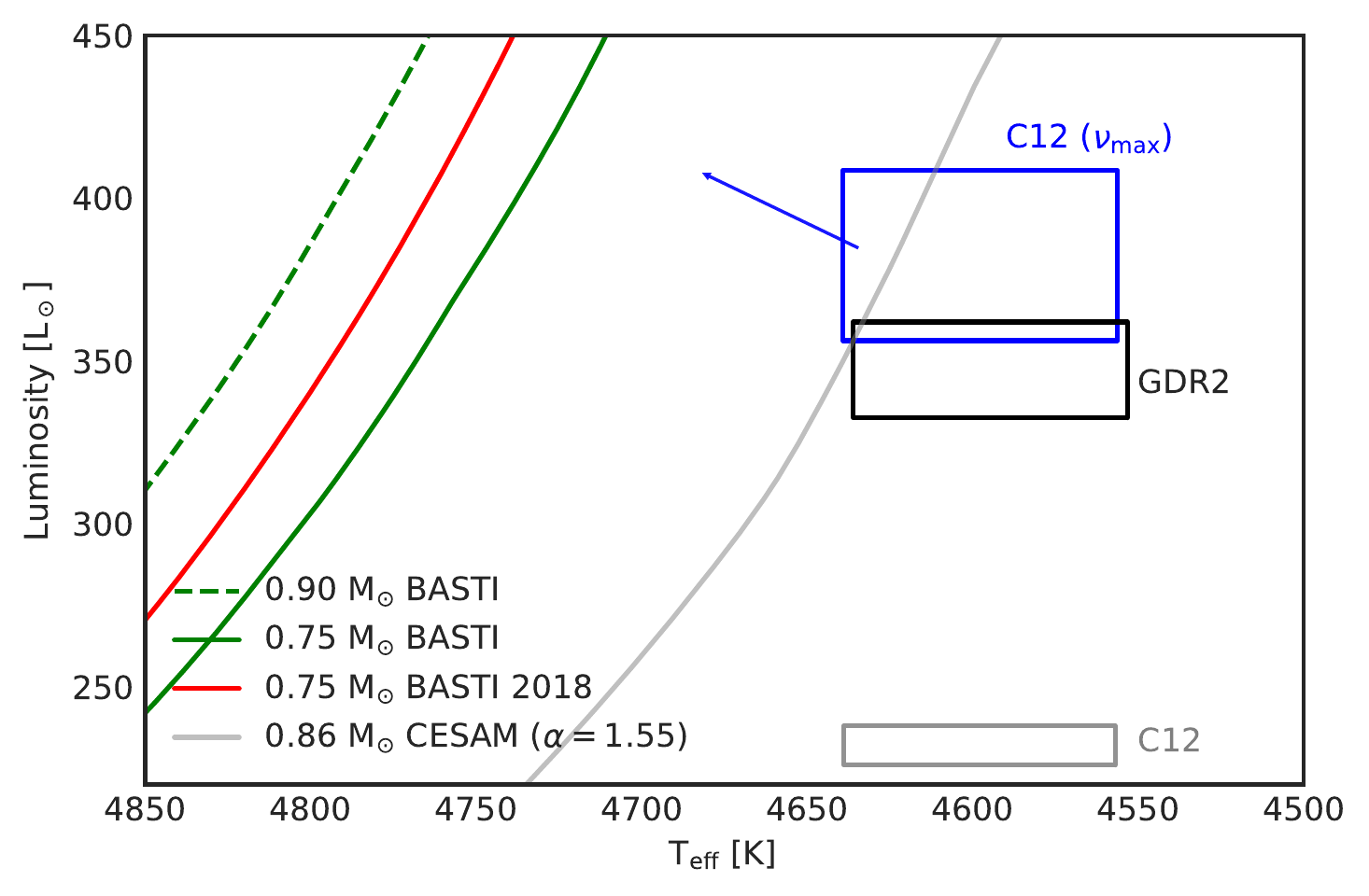}%
  \caption{{\bf Left:} Density plots of derived distance versus \logg\ using asteroseismology --- without (blue) and with (green) corrections to the seismic relation --- and using Gaia DR2 (black)  {\bf Right:} HR diagram showing observational position of HD\,122563 using different input data (blue, black, and grey), along with standard evolutionary models from BASTI and a tailored CESAM model.}
  \label{Creevey:fig2}
\end{figure}

We derived $d$ and \logg\ using the same methodology as above, but by using the set 
($\pi, \theta$) and the mass prior, where $\pi$ is the parallax from Gaia DR2 \citep{gaia2018a}.
In Fig.~\ref{Creevey:fig2} the solution is represented by the black contours (left) and the black box (right).  

As can be seen in both figures, similar solutions are obtained using the independent approaches. \logg\ differs only by 0.02 -- 0.04 dex, and $d$ by less than 1.5$\sigma$. Consequently the \lum\ agree also to 1$\sigma$ (1$\sigma$ error boxes are shown on the right panel).  
In Fig~\ref{Creevey:fig1} right panel we also show standard evolutionary tracks from the 
BASTI models in green \citep{basti04}, and a model from the updated tracks in red \citep{basti18} using solar-scaled canonical models\footnote{$\alpha$-enhanced tracks from \cite{basti04} are hotter, and those from \cite{basti18} are not yet available.}.
Using fine-tuned evolution models from the CESAM evolutionary code \citep{morel97}, we could
reproduce the observed position but only be reducing the mixing-length parameter by 0.3 from the reference solar value.

\section{Conclusions\label{sec:conclusions}}
\begin{itemize}

\item We have detected oscillations in the metal-poor star HD\,122563.
\item By comparing the distances derived using asteroseismology and that from a parallax from Gaia, we showed that the scaling relations for surface gravity work in the metal-poor and evolved regime.
\item We have derived new fundamental parameters for this star using asteroseismology: $\log g$ = 1.39 $\pm$ 0.01 and $d$ = 306 $\pm$ 9 pc ($d_{\rm GDR2} = 290 \pm 5$ pc).
\item By applying corrections to the scaling relations for molecular weight and the adiabatic exponent we derived values of $\log g = 1.42 \pm 0.01$ and $d = 296 \pm 9$ pc.
\item The new fundamental parameters imply less tension between evolution models although a discrepancy on the order of 150-200 K still exists.  This can be remedied by reducing the mixing-length parameter by 0.3 compared to solar, but further studies with more realistic physics in these models should also be addressed.
\item Increasing $A_V$ could also alleviate some of the problem.
\item The new surface gravity implies less tension with 3D models, see e.g. \citet{collet18} who suggest that a lowering of \logg\ in their analysis to alleviate discrepancies between their molecular- and atomic-species-derived oxygen abundances.
\item We continue to collect data from the SONG  Hertzsprung telescope.  We aim to detect the mean frequency separation, along with individual frequencies, and a more accurate determination of the width of \numax\ (see \citet{yu2018} who indicate a trend of width versus \numax).
\item We also aim to understand the long-term trends seen in the time series, see Fig.~\ref{Creevey:fig1}
\item The individual frequencies will be very instructive for improving the theoretical models in
the metal-poor regime.

\end{itemize} 
\begin{acknowledgements}
We are grateful for the Programme National de Physique Stellaire for financial support for this research project. This work is based on observations made with the Hertzsprung SONG telescope operated on the Spanish Observatorio del Teide on the island of Tenerife by the Aarhus and Copenhagen Universities and by the Instituto de Astrof\'isica de Canarias. Support for the construction of the Hertzsprung SONG Telescope from the Instituto
de Astrofisica de Canarias, the Villum Foundation, the Carlsberg Foundation and
the Independent Research Fund Denmark is gratefully acknowledged.  E.C. is funded by the European Union's Horizon 2020 research and innovation program under the Marie Sklodowska-Curie grant agreement No. 664931.
We also deeply acknowledge the Gaia Data Processing and Analysis Consortium for their huge efforts in bringing us high quality data.

\end{acknowledgements}

\bibliographystyle{aa}  
\bibliography{Creevey_S15} 

\end{document}